\documentclass{article}

\usepackage{arxiv}

\usepackage[utf8]{inputenc} 
\usepackage[T1]{fontenc}    
\usepackage{hyperref}       
\usepackage{url}            
\usepackage{booktabs}       
\usepackage{amsfonts}       
\usepackage{nicefrac}       
\usepackage{microtype}      
\usepackage{lipsum}
\usepackage{graphicx}
\usepackage{siunitx}
\usepackage{placeins}
\graphicspath{ {./images/} }
\usepackage{hyperref}
\usepackage{breakurl}
\usepackage{siunitx}
\usepackage{subcaption}
\usepackage{doi}
\usepackage{bm}

\title{BINAQUAL: A Full-Reference Objective Localization Similarity Metric for Binaural Audio}

\author{
 Davoud Shariat Panah \\
  School of Computer Science\\
  University College Dublin\\
  Dublin, Ireland \\
  \texttt{davoud.shariatpanah@ucd.ie} \\
   \And
  Dan Barry\\
	School of Computer Science\\
	University College Dublin\\
	Dublin, Ireland\\
	\texttt{danbarry@duck.com} \\
  \And
	Alessandro Ragano\\
	School of Computer Science\\
	University College Dublin\\
	Dublin, Ireland\\
	\texttt{alessandro.ragano@ucd.ie} \\
    \And
	Jan Skoglund\\
	Google LLC\\
        San Francisco, CA\\
	\texttt{jks@google.com} \\
    \And
	Andrew Hines\\
	School of Computer Science\\
	University College Dublin\\
	Dublin, Ireland\\
	\texttt{andrew.hines@ucd.ie} \\
}

\begin{document}
\maketitle
\begin{abstract}
Spatial audio enhances immersion in applications such as virtual reality, augmented reality, gaming, and cinema by creating a three-dimensional auditory experience. Ensuring the spatial fidelity of binaural audio is crucial, given that processes such as compression, encoding, or transmission can alter localization cues. While subjective tests are used for evaluating spatial localization quality, they are costly and time-consuming. This paper introduces \mbox{BINAQUAL}, a full-reference objective metric designed to assess localization similarity in binaural audio recordings. \mbox{BINAQUAL} adapts the \mbox{AMBIQUAL} metric, originally developed for localization quality assessment in ambisonics audio format to the binaural domain. We evaluate \mbox{BINAQUAL} across five key research questions, examining its sensitivity to variations in sound source locations, angle interpolations, surround loudspeaker layouts, audio degradations, and content diversity. Results demonstrate that \mbox{BINAQUAL} effectively differentiates between subtle spatial variations and correlates strongly with subjective listening tests, making it a reliable metric for binaural localization quality assessment. The proposed metric provides a robust benchmark for ensuring spatial accuracy in binaural audio processing, paving the way for improved objective evaluations in immersive audio applications.
\end{abstract}

\section{Introduction}\label{sec:intro}
Spatial audio is an advanced auditory technology that creates a three-dimensional sound field, enabling listeners to perceive sounds as originating from specific locations in space. This technology enhances user immersion across applications such as virtual reality (VR), augmented reality (AR), gaming, and cinematic soundscapes. Binaural audio, a form of spatial audio, utilizes two channels to simulate the physical and perceptual cues involved in sound localization. With smartphones, headphones, and wearable devices emerging as dominant platforms for music, movies, and interactive media, there is a growing interest in immersive auditory experiences that can be delivered in a convenient and effective manner \cite{Pangarkar2025Smart}. In this context, binaural audio is expected to see significant growth, driven by evolving media consumption patterns.

The rise of open-source spatial audio formats, such as Eclipsa Audio \cite{eclipsa_audio}, alongside binaural remixing of content for headphone playback, is reshaping the way immersive sound is created and experienced. Such innovations highlight the growing need for robust localization quality metrics to assess and optimize spatial accuracy in binaural audio rendering. As content creators and engineers increasingly adopt object-based audio mixing and dynamic head-tracked rendering, existing metrics often fail to reflect perceptual accuracy across diverse playback environments. Developing quantitative and perceptually aligned localization quality metrics is crucial for ensuring consistent auditory immersion in VR, gaming, cinema, and streaming services, where precise spatial positioning enhances realism and listener engagement. 

 Localization quality assessment refers to the process of quantifying the accuracy and perceptual fidelity of spatial audio reproduction, typically by comparing perceived sound source locations with their intended positions. While MUSHRA-style protocols \cite{RECOMMENDATION} are occasionally adapted for evaluating spatial attributes, localization accuracy is more commonly assessed using task-specific methods that directly measure directional perception, such as directional pointing tasks (e.g., Makous et al. \cite{Makous1990Twodimensional}) or Minimum Audible Angle thresholds \cite{mills1958minimum}. However, such tests are time-consuming, expensive, and prone to variability due to individual listener differences and environmental factors \cite{Bech2007Perceptual}. As the demand for reliable and reproducible evaluations grows, effective objective metrics are needed to assess spatial localization quality in diverse contexts. 
 
 Developing effective localization quality metrics is a challenging task. Factors such as variabilities in individual listeners, recording environments, and stimulus characteristics can significantly influence the spatial cues in binaural recordings \cite{Ihlefeld2011Effect}. In recent years, a variety of spatial localization quality metrics have been proposed. Such metrics can be broadly categorized into signal processing-based techniques \cite{Choi2007Prediction,Flessner2017Assessment,Flener2019Subjective,Watcharasupat2023Quantifying}, and data-driven models \cite{Flesner2014Quality,Manocha2021DPLM,Manocha2022SAQAM,Manocha2023Spatialization}. However, the robustness of the metrics to variations in binaural recordings remains poorly understood, primarily due to the lack of rigorous benchmarks. Also, the reproducibility of these metrics is limited, as their source codes are often unpublished. 

To address these gaps, this paper introduces a full-reference objective metric called \mbox{BINAQUAL}, designed to assess spatial localization similarity in binaural audio recordings. Given a binaural audio signal and its corresponding reference, \mbox{BINAQUAL} provides a score that measures how perceptually similar their spatial localization characteristics are. \mbox{BINAQUAL} is developed by adapting an objective localization quality metric originally designed for ambisonic audio signals \cite{Narbutt2020AMBIQUAL}. Through extensive experiments and comprehensive benchmarking, we investigate the robustness of the proposed metric with respect to environmental and stimulus-related factors. To achieve this, we generate multiple synthetic datasets tailored to the research objectives of the study. As a key contribution of this work, we compile and publicly release a combined dataset, along with the source code of our metric, to facilitate further research and ensure reproducibility.

To rigorously evaluate the \mbox{BINAQUAL}, we first examine its sensitivity to small variations in the spatial locations of point sources. If the metric fails to distinguish between sounds originating from different locations, it may instead be responding to differences in fidelity, intensity, or other confounding factors rather than localization cues. Understanding this sensitivity is fundamental to ensuring the metric's validity for spatial localization evaluation.

Second, we examine whether the metric can differentiate between binaural renders that use real versus interpolated impulse responses. Ideally, binaural rendering systems would have access to real impulse responses for all source angles, but in practice, interpolation is often used to approximate missing data. Evaluating the ability of the model to distinguish between real angle versus angle interpolation renders ensures that it is capable of assessing the accuracy of binaural renderings.

Third, we investigate whether the metric can distinguish between simulated binaural representations of different surround loudspeaker layouts. Many spatial audio recordings are mixed specifically for distinct surround layouts, such as 5.1 or 7.1. When down-mixed for binaural listening, preserving spatial cues is essential. A metric capable of distinguishing between these layouts ensures that the spatial localization of the audio is accurately preserved during the binaural rendering of multichannel surround-encoded audio.

Fourth, we evaluate the metric's ability to distinguish between different levels of codec compression applied to binaural audio recordings. Compression algorithms can introduce distortions that impact spatial cues, hence a reliable localization quality metric should be able to capture these degradations.

Real-world audio scenes often contain complex spatial distributions of multiple sources, making it essential for the metric to perform reliably under diverse conditions. Therefore, we assess the robustness of the metric across different types of content and varying numbers of point sources to ensure its reliability and practicality.

To address the research objectives discussed above, we pose the following research questions:

\begin{itemize}
    \item \textbf{RQ1:} Is the proposed metric sensitive to variations in the spatial locations of the point sources?
    \item \textbf{RQ2:} Can the metric differentiate between single-point source binaural renders that use real versus interpolated angle impulse responses?
    \item \textbf{RQ3:} Can the metric distinguish between the simulated binaural representations of different surround loudspeaker layouts?
    \item \textbf{RQ4:} Can the metric distinguish between different codec compression levels of binaural audio recordings?
    \item \textbf{RQ5:} How robust is the metric to variations in content and number of point sources in binaural audio recordings?
\end{itemize}

While several localization quality metrics have been proposed in previous work, they primarily focus on accuracy prediction without addressing factors affecting localization robustness. To advance spatial audio quality research beyond simple accuracy estimation, we investigate key research questions that have not been explicitly addressed in prior work. Addressing these questions allows us to establish the robustness, reliability, and practical utility of the \mbox{BINAQUAL} across diverse scenarios. A metric that can reliably evaluate localization quality under these conditions will be more applicable across a broad range of use cases, ensuring that it is practical for real-world deployment. By addressing the above research questions, the \mbox{BINAQUAL} metric sets a benchmark for assessing localization quality in binaural audio with accuracy, reliability, and versatility. 

The remainder of this paper is organized as follows. Sec. \ref{sec:background} provides the background on binaural audio and sound source localization, along with a review of existing spatial localization quality metrics. Additionally, this section includes a summary of the \mbox{AMBIQUAL} metric, which has been adapted in this study. Sec. \ref{sec:binaqual} introduces the \mbox{BINAQUAL} model for assessing spatial localization similarity in binaural audio signals. Sec. \ref{sec:datasets} describes the evaluation datasets used in the study. In Sec. \ref{sec:results}, the results are presented and discussed. Finally, Sec. \ref{sec:conclusion} concludes the paper and outlines potential future directions.

\section{Background}\label{sec:background}
\subsection{Binaural Audio}
Binaural audio is an audio format designed to replicate the way humans naturally perceive sound, creating a realistic and immersive auditory experience. Unlike stereo audio, which typically focuses on directional cues in a flat plane, binaural audio simulates the three-dimensional sound field by incorporating spatial cues \cite{Mller1992Fundamentals}. This allows listeners to perceive sounds as originating from specific locations in space as if they were physically present in the sound environment. 

The perceived authenticity of binaural audio relies on the accurate simulation of spatial auditory cues that humans use to localize sound sources. Among these, interaural time differences (ITD) and interaural level differences (ILD) play key roles in lateral localization. ITD refers to the small delay in the arrival time of a sound at one ear compared to the other, and is most effective for low-frequency sounds. ILD describes the difference in sound intensity between the ears, primarily due to the head shadowing effect, and is more prominent for high-frequency sounds \cite{Moore2010Oxford, Macpherson2002Listener}. However, these are not the only cues involved in spatial perception. For a plausible and perceptually authentic binaural experience, additional factors such as spectral shaping by the outer ear and dynamic cues related to head movement also contribute significantly to sound source localization in elevation and distance \cite{jc1991sound, Blauert1997Spatial}

Binaural audio can be created through recording or synthesis. Binaural recording involves capturing audio using specially designed microphones placed inside the ear canals of a real or artificial head. However, binaural recordings are inherently static; once recorded, the spatial cues are fixed, preventing listeners from interacting with the soundscape by moving their heads or changing their positions \cite{Begault20003}. Additionally, the characteristics of the head used in the recording process may not match those of individual listeners, leading to inaccuracies in spatial perception \cite{Mller1992Fundamentals}. As a result, this technique has limited applicability in dynamic or personalized real-world scenarios.

In the binaural synthesis technique, spatial audio is synthesized by processing sound signals through head-related transfer functions (HRTFs). The HRTF describes how an individual’s body, particularly the head, torso, and outer ears filter sound as it travels from a source to the eardrum \cite{Mller1992Fundamentals}. By applying the appropriate HRTFs based on the intended location of each sound source, it is possible to create a binaural representation of a complex sound field. Binaural synthesis allows for dynamic and customizable spatial audio experiences but depends on the accuracy and availability of high-quality HRTF datasets \cite{Begault20003}. For the experiments in this study, we use the SADIE II database \cite{Armstrong2018Perceptual} to synthesize binaural audio samples. This database offers a comprehensive collection of HRTFs that cover a wide range of azimuths and elevations, enabling accurate spatial representation of sound sources.

\subsection{Sound Source Localization}\label{sec:ssl}
Sound source localization is the process of identifying the spatial location of a sound source, including its azimuth, elevation, and distance, based on auditory cues extracted from the sound waves received by a hearing system or sensor array \cite{Blauert1997Spatial}. This process leverages differences in time, intensity, and frequency content between signals captured at multiple points, such as ears or microphones, to determine the source's location in a three-dimensional space.

Multiple factors influence the accuracy of sound source localization in humans. The first factor is individual listener characteristics such as age, head and ear anatomy, and prior auditory experience. It has been shown that age negatively impacts the accuracy of sound source localization in older adults in terms of both azimuth and elevation \cite{dobreva2011influence}. Individual differences in head size, ear shape, and torso affect HRTFs, which describe how sounds are filtered by the body before reaching the eardrum. These variations influence localization accuracy \cite{Hofman1998Relearning, Brungart1999Auditory}. In addition to these, prior auditory experience, such as familiarity with a sound or training in spatial hearing, enhances the ability to interpret localization cues \cite{Wightman1999Resolution}.

Characteristics of the environment are another factor that influences localization accuracy in humans. Reverberation occurs when sound waves reflect off surfaces, creating multiple delayed copies of the original sound. Reverberant environments blur the ITD, ILD, and spectral cues, which result in larger errors in sound source localization \cite{Hartmann1983Localization}. The presence of background noise in the environment can also reduce localization accuracy by masking the target sound and reducing the ability to extract spatial cues \cite{Kerber2012Sound,Ege2018Accuracy}.

Lastly, the characteristics of the stimulus also significantly influence the localization accuracy. Makous et al. \cite{Makous1990Twodimensional} investigated the influence of sound location on localization error in humans. They found that sound localization is very accurate for sources located directly in front of the listeners, where the average error was 2.0° and 3.5° for horizontal and vertical dimensions, respectively. However, as the sound sources were shifted towards peripheral locations, the errors increased significantly, reaching up to 20° at the extreme positions. The impact of the spectral content of the stimulus on the accuracy of sound source localization has also been investigated in several studies. It has been shown that both frequency content and bandwidth of the signal affect the accuracy of localization \cite{Shigeno1983Localization,Chau1995Combined,Borg2008Effect}. High-frequency spectral cues improve elevation judgments and help resolve front-back confusion \cite{Middlebrooks1992Narrow,Butler1992Localization}. Localization accuracy for low-frequency sounds is typically good in azimuth but poorer in elevation due to the lack of spectral cues \cite{Middlebrooks1992Narrow,Wightman1992dominant}. Regarding the sound bandwidth impact, previous work showed that wider stimulus bandwidth leads to higher localization accuracy and vice versa \cite{Chau1995Combined,Borg2008Effect,Yost2014Sound}. This is because signals encompassing a wider range of frequencies provide richer spatial cues, leading to improved localization accuracy.

A wide range of sound source localization models have been proposed \cite{zakarauskas1993computational, langendijk2002contribution, dietz2011auditory, may2010probabilistic}. Some of these models, such as those by Lindemann and Baumgartner, are available in the MATLAB Auditory Modeling Toolbox \cite{lindemann1986extension, baumgartner2014modeling}.

\subsection{Spatial Localization Quality Metrics}
Significant efforts have been made to develop objective metrics for assessing localization quality in spatial audio content. The work in this domain can be broadly categorized into signal processing-based and data-driven techniques.

Choi et al. \cite{Choi2007Prediction} proposed a model that leverages ILD, ITD, and interaural cross-correlation coefficient (IACC) distortions to measure perceived quality in multi-channel audio. Flessner et al. \cite{Flessner2017Assessment} introduced BAM-Q, a metric that uses a binaural auditory model as a front-end to extract ILD, ITD, and interaural coherence from binaural signals, which are then utilized for spatial quality assessment. Joint modeling of monaural and binaural signal distortions has also been explored. For example, the work in \cite{Flener2019Subjective} combines BAM-Q with a generalized power-spectrum model to predict quality based on both types of distortions. More recently, Watcharasupat et al. \cite{Watcharasupat2023Quantifying} presented a spatial quality evaluation model using a filter decomposition technique based on the duplex theory of spatial hearing. In another recent work, Ren et al. \cite{ren2025metric} proposed an objective metric called KSRIR to predict the listening quality and localization accuracy of first-order ambisonics generated using interpolated room impulse responses. The quality scores were computed by comparing the cumulative distribution functions (CDFs) of the reference and synthesized signals using the Kolmogorov-Smirnov test.

An early contribution to data-driven approaches was made by Flessner et al. \cite{Flesner2014Quality}, who proposed a spatial quality assessment model based on multivariate adaptive regression splines (MARS). Manocha et al. \cite{Manocha2021DPLM} introduced DPLM, a full-reference spatial localization quality metric based on a direction-of-arrival (DOA) deep network model. They extended this work with SAQAM \cite{Manocha2022SAQAM}, a multi-task learning model that evaluates both spatial localization quality and listening quality of binaural signals. However, these models were limited to audio signals with a 16 kHz sampling rate. To address this limitation, the authors recently proposed a new spatial quality assessment metric for binaural speech \cite{Manocha2023Spatialization}. This metric leveraged a multi-task learning framework with direction-of-arrival estimation and binaural speech synthesis as auxiliary tasks. While this model is sampling rate agnostic, its source code is not available, which limits the reproducibility of the work. More recently, Zheng et al. \cite{zheng2025hapg} proposed a spatial quality metric called HAPG-SAQAM for evaluating the spatial and overall audio quality of binaural signals. Their method uses gammatone frequency cepstral coefficients along with perceptual weighted loss, showing improvements over the SAQAM metric.

\begin{figure*}[h]
\centering
\includegraphics[width=\textwidth]{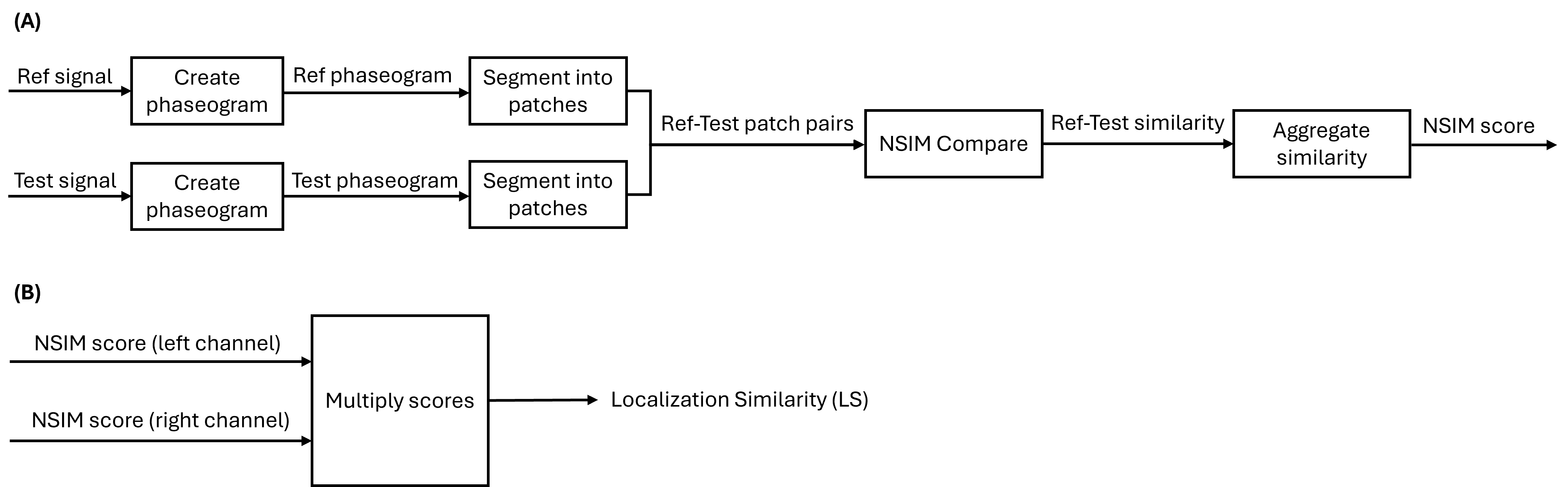}
\caption{A high-level representation of the \mbox{BINAQUAL} model. (A) The process to calculate NSIM scores for each channel of reference and test signals. (B) The process to calculate localization similarity from NSIM scores.}
\label{binaqual_model}
\end{figure*}

\subsection{AMBIQUAL}
AMBIQUAL \cite{Narbutt2020AMBIQUAL} is a full-reference objective metric designed to assess the quality of compressed ambisonic spatial audio. It evaluates both listening quality and localization accuracy directly from B-format ambisonic. More specifically, listening quality is determined from the omnidirectional channel, while localization accuracy is calculated based on a weighted product of similarities from the directional channels in the B-format audio.

AMBIQUAL was developed by adapting and extending ViSQOL \cite{Hines2015ViSQOLAudio}, a full-reference objective metric designed to evaluate the perceptual quality of audio signals. ViSQOL operates by comparing a reference (high-quality) audio signal with a test (degraded) version. For this purpose, it uses the Neurogram Similarity Index Measure (NSIM) \cite{Hines2012Speech}, which measures the similarity between corresponding time-frequency components of the reference and test signals, providing a quantitative indication of how well the test signal preserves the spectral characteristics of the reference.

Unlike ViSQOL, which uses magnitude spectrograms to measure similarities between reference and test signals, \mbox{AMBIQUAL} leverages spectrograms of phase angles (phaseograms). Phaseograms are segmented into fixed-duration patches, and NSIM is applied locally to each aligned reference and test patch. The final similarity score is obtained by averaging NSIM values across all patches and frequency bands. These scores are aggregated across directional channels to estimate localization fidelity.

AMBIQUAL quality predictions were compared with the results of a MUSHRA subjective listening test for a varied set of compressed ambisonics contents, and it was shown that \mbox{AMBIQUAL} predictions correlate well with scores from the listening test \cite{Narbutt2020AMBIQUAL}. However, the \mbox{AMBIQUAL} model is limited to ambisonic content, as it operates on B-format rather than the binaural render. In other words, the model provides a single quality score for an ambisonic recording, regardless of the render angle.

\section{BINAQUAL MODEL}\label{sec:binaqual}
\mbox{BINAQUAL} is a full-reference objective metric for assessing spatial localization similarity in binaural audio recordings. \mbox{BINAQUAL} is designed by adapting the \mbox{AMBIQUAL} model to two-channel binaural audio signals. In practice, \mbox{BINAQUAL} could be applied to an ambisonic signal, but only for a particular render. Our hypothesis was that adapting the \mbox{AMBIQUAL} model and applying it at the post-render stage could allow any binaurally rendered spatial audio signals to be evaluated, regardless of their origin mix or format.

As mentioned earlier, in \mbox{AMBIQUAL}, localization accuracy is calculated using a weighted product of similarities from directional channels of B-format ambisonics. Given that binaural audio signals consist of only two channels (left and right), \mbox{BINAQUAL} calculates localization similarity as the product of similarities between the reference and test signals in these two channels. We should note that \mbox{BINAQUAL} is a relative localization similarity measure, not an absolute localization error estimator (i.e., angular degrees).

Fig.~\ref{binaqual_model} illustrates a high-level representation of the \mbox{BINAQUAL} model. As shown in part (A), the process for calculating the NSIM score for reference and test signals resembles the approach used in the \mbox{AMBIQUAL} model. A summary of this process is provided below, and readers are encouraged to refer to \cite{Narbutt2020AMBIQUAL} for additional details.


As shown in part (A) of Fig.~\ref{binaqual_model}, spectrograms of phase angles (phaseograms) are created for each channel of the reference and test signals. These phaseograms are computed using a 2048-point Short-Time Fourier Transform (STFT) with a 1536-point Hamming window and 50\% overlap. To align the analysis with the perceptually-motivated frequency range of our 32-band gammatone filterbank (50 Hz to 16 kHz), a specific subset of STFT frequency bins is selected. We use the 640 bins corresponding to indices 2 through 641 of the original STFT. At a 48 kHz sampling rate, this selection captures a frequency range from approximately 47 Hz to 15.02 kHz. This range effectively covers the bands of interest from the gammatone filterbank while excluding the DC component (bin 0).

NSIM is then used to compute localized similarity between patch pairs using two components: an intensity term that reflects the similarity of local means, and a structure term that captures the correlation (covariance) between patches. Both components are stabilized with constants derived from the dynamic range, which in the case of phaseograms corresponds to $ 2\pi$. These frame-wise similarity scores are computed over all 640 frequency bins and then averaged within 32 perceptually motivated gammatone frequency bands. The band-averaged scores are further aggregated (averaged) across time patches to produce a single global similarity score per channel. The final NSIM score ranges from 0 to 1, where 1 indicates identical patches and values closer to 0 indicate low similarity. This bounded nature makes NSIM particularly well-suited for perceptual similarity tasks, offering normalized and interpretable scores, unlike unbounded metrics such as RMSE. Readers are encouraged to refer to \cite{Hines2012Speech} for further details about NSIM.

This process is repeated for both channels of the reference and test signals. As shown in part (B) of Fig.~\ref{binaqual_model}, the localization similarity score is calculated as the product of the final NSIM scores from the two channels.

The motivation for applying NSIM to phaseograms of binaural audio signals lies in the relationship between phase information and the spatial cues essential for sound localization. As mentioned earlier, ITDs dominate low-frequency localization, while ILDs dominate at high frequencies. These cues are encoded in the temporal and spectral structure of binaural signals. By comparing the phaseograms of left and right channels between reference and test signals, \mbox{BINAQUAL} indirectly assesses the fidelity of the spatial cues, particularly ITD and ILD, embedded in those signals. Since the NSIM metric is sensitive to both local phase distortions and structural mismatches across frequency bands, it provides a perceptually relevant measure of how well the test signal preserves spatial localization characteristics. Thus, high similarity between binaural phaseograms implies preserved localization cues, whereas reduced similarity suggests perceptual degradation in spatial fidelity.

The source code of the \mbox{BINAQUAL} model is available online\footnote{\url{https://github.com/QxLabIreland/Binaqual}}.

\section{EVALUATION DATASETS}\label{sec:datasets}
To present a comprehensive evaluation of the BINAUQAL model, three evaluation datasets were created. These datasets were generated synthetically by applying specific HRTFs from subject D2 (dummy head) of the SADIE II database \cite{Armstrong2018Perceptual} to various audio contents using the Binamix Python library \cite{binamix}. When the desired angle was available in the SADIE database, the corresponding binaural impulse response (BIR) was used directly. For angles not present in the database, suitable BIRs were generated via interpolation at render time. This was achieved by identifying the nearest available angles using a modified Delaunay triangulation and computing interpolation weights based on Euclidean distances. For further details on the interpolation method, readers are referred to \cite{binamix}.

The datasets are specifically designed to address research questions one, two, and three, as outlined in Sec. \ref{sec:intro}. To address research question four, we use the same dataset used in \cite{Narbutt2020AMBIQUAL}. It is worth mentioning that we do not create a separate dataset for research question five; instead, the robustness of the model to various contents and the number of point sources is assessed within the scope of research questions one to four. The compiled dataset is called SynBAD (Synthetic Binaural Audio Dataset) and is available online\footnote{\url{https://zenodo.org/records/15431990}}. A detailed description of each dataset is provided below.

\subsection{Localization Sensitivity Dataset}
The localization sensitivity dataset is designed to evaluate the model's sensitivity to the locations of point sources (RQ1). To generate the dataset, a subset of HRTFs with varying azimuths and elevations from the SADIE II database was selected. The selection was based on the average human performance in sound source localization. As mentioned in Sec. \ref{sec:ssl}, human localization errors tend to increase as sound sources shift toward peripheral locations. To account for this, the average unsigned localization errors reported in \cite{Makous1990Twodimensional} across different azimuth and elevation ranges were used to determine a subset of HRTFs to synthesize binaural samples. Tables \ref{table:azimuth_increments} and \ref{table:elevation_increments} summarize the increments applied for different azimuth and elevation ranges, respectively.

To evaluate the robustness of the model to different contents, ten mono audio samples were used to synthesize the dataset. As summarized in Table \ref{table:audio_content}, these audio samples were either synthetically generated or originated from the EBU dataset \cite{Tech19883253}, and cover a wide range of frequencies and bandwidths. Three samples with reverberation effect were also included in the base sample set to analyze the model's performance in reverberant conditions.

Using the angles and audio content described above, 14,500 binaural samples were synthesized at a sampling rate of 48 kHz. Since the dataset was intended solely for evaluation and not for subjective listening tests, each sample was limited to a duration of 2 s, sufficient for assessing model sensitivity to point source locations.

\subsection{Angle Interpolations Dataset}


The angle interpolations dataset is designed to evaluate the model's ability to differentiate between point source renders using discrete angles versus interpolated angles (RQ2). To create this dataset, we binaurally rendered the audio contents described in Table \ref{table:audio_content} using various degrees of interpolation between angle pairs. These angle pairs have been summarized in Table \ref{table:rq3}. As shown in Table \ref{table:rq3}, the first set of angles is used to render a single source virtually at azimuth, elevation = (50°, 0°) while the second set is used for a virtual render at (90°, 0°). As an example, a reference at (50°, 0°)  can be virtually represented by interpolation between any of the following angle pairs: \{(45°, 0°) and (55°, 0°)\}, \{(40°, 0°) and (60°, 0°)\}, \{(30°, 0°) and (70°, 0°)\}, \{(20°, 0°) and (80°, 0°)\}, \{(10°, 0°) and (90°, 0°)\}, \{(0°, 0°) and (100°, 0°)\}.

Using the angle pairs described in Table \ref{table:rq3}, 140 binaural samples with a duration of 5~s and sampling frequency of 48 kHz were synthesized.

\subsection{Surround Layouts Dataset}
The surround layouts dataset is designed to evaluate the model's ability to distinguish between simulated binaural representations of various surround loudspeaker layouts, including 5.1, 5.1.4, 7.1, and 7.1.4 (RQ3). The previous datasets included sources rendered at a large variety of angles. However, the surround layouts dataset contains renders of various contents at only two sets of angles, one with elevation and one without. This is because we will only use this dataset to investigate the sensitivity of the metric to the presence of elevation (i.e., height loudspeaker) in surround loudspeaker setups.

To generate samples of this dataset, for each of the surround layouts mentioned above, two point sources were rendered at the following angles (azimuth, elevation): 
(60°, 0°) and (300°, 0°), as well as (60°, 50°) and (300°, 30°). We also rendered the sources at the above angles without any loudspeaker layouts to be used as reference signals. 

The Binamix library \cite{binamix} was used to handle the rendering of sources for specific surround loudspeaker layouts. This library uses impulse response interpolation to simulate the VBAP method \cite{pulkki1997virtual}, rendering the source at the virtual loudspeaker positions between the discrete angles available for a specific loudspeaker layout. The audio contents described in Table \ref{table:audio_content} were mixed in different ways to ensure a diverse set of sound source combinations in this dataset. These combinations, along with the purpose of each, have been detailed in Table \ref{table:rq2}.

Using the angles and audio contents described above, 80 binaural samples with a duration of 5~s and sampling frequency of 48 kHz were synthesized.

\subsection{Codec Compression Dataset}
The codec compression dataset is employed to test the model’s ability to distinguish between different compression levels applied to binaural recordings (RQ4). This dataset contains first-order (FOA) and third-order (3OA) ambisonics of various audio contents. The samples include both single-point and two-point audio sources with a variety of localizations, including dynamic and fixed sources. The audio samples were encoded using the Opus 1.2 codec, channel mapping family 2 at various bit rates, including 512, 384, 256, 128, 96, 64, and 32 kbps to produce a range of conditions. They were then rendered binaurally for presentation. Readers are encouraged to refer to \cite{Narbutt2020AMBIQUAL} for a full description of the dataset samples.

\begin{figure*}[h]
\centering
\includegraphics[width=\textwidth]{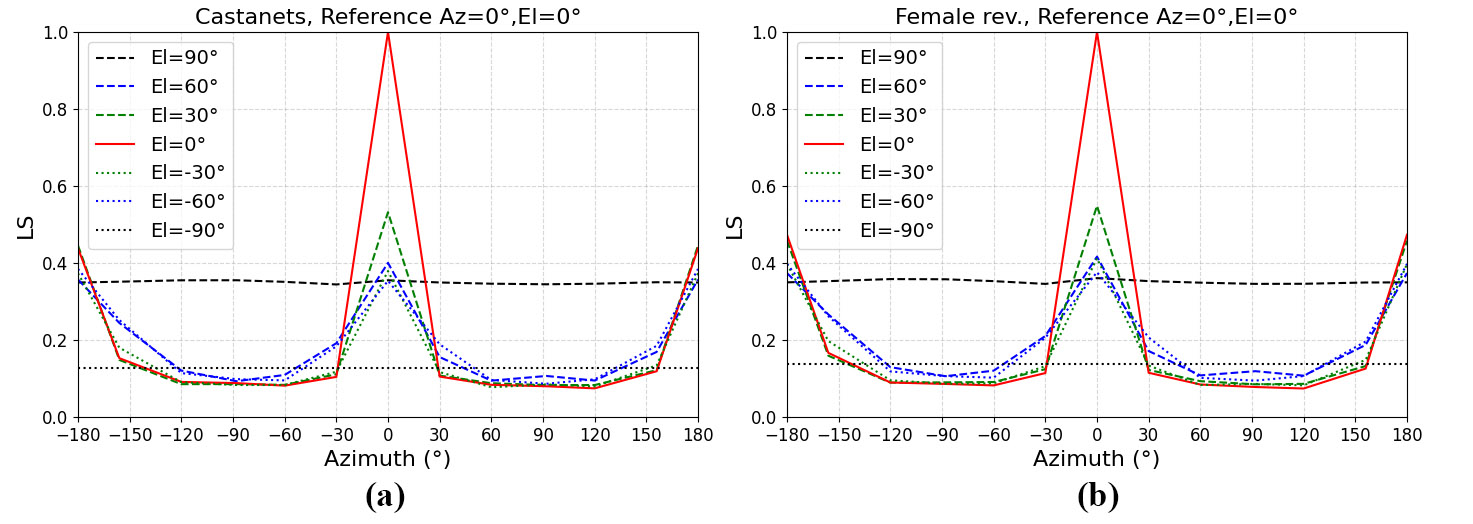}
\caption{Localization similarity as a function of azimuth and elevation with fixed reference audio source, localized at azimuth=0°, elevation=0° for (a) Castanets and (b) Female speech w. reverb test signals.}
\label{RQ1_line}
\end{figure*}

\section{RESULTS AND DISCUSSION}\label{sec:results}
This section presents and analyzes the results with respect to five research questions outlined in Sec. \ref{sec:intro}. 

\subsection{Localization Sensitivity}\label{sec:results_ls}
Localization similarity (LS) was computed for each sample in the localization sensitivity dataset, using a reference source localized at azimuth = 0°, elevation = 0°. Fig.~\ref{RQ1_line} illustrates the localization similarity as a function of azimuth and elevation angles for female and castanets sources.

As shown in Fig.~\ref{RQ1_line}, for all elevations except ±90°, the localization similarity between test and reference signals decreases as the test azimuth moves further away from the reference. However, an increase in LS is observed between -120° to -180° and 120° to 180°, which can be attributed to front-back confusion. For elevations of ±90°, the LS curve remains relatively flat across azimuths. This is expected, as at ±90° elevation, corresponding to directly above or below the listener, azimuth resolution effectively collapses. In other words, in Cartesian coordinates, all azimuths converge to a single point, making azimuth distinctions meaningless. Additionally, binaural localization cues such as ITD and ILD become minimal at these elevations, and although spectral cues are still present, they offer limited azimuthal discrimination, further contributing to a uniform LS profile.

\begin{figure*}[!t]
\centering
\includegraphics[width=\textwidth]{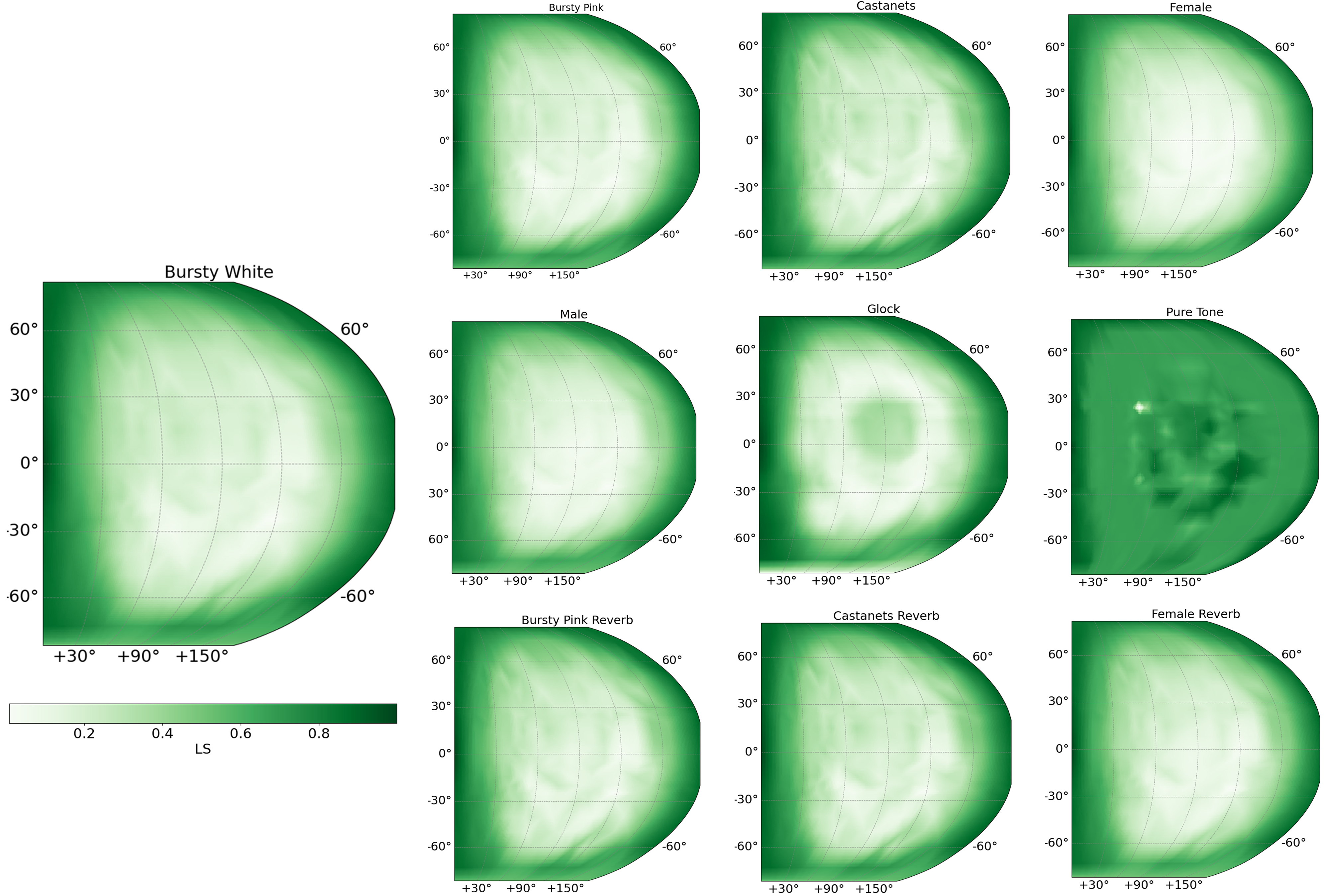}
\caption{Localization similarity scores distributed on a half sphere for various contents. The corresponding reference audio sources were localized at azimuth = 0°, elevation = 0°. Box-Cox transform was applied to LS scores to spread out the distribution and make variations more noticeable.}
\label{RQ1_sphere}
\end{figure*}

Fig.~\ref{RQ1_sphere} shows half-hemisphere heatmaps of LS for all contents in the localization sensitivity dataset. To create these heatmaps, we used Robison projection from Cartopy, a Python library for geographic map generation. Since the LS scores were in a small range (mostly between 0--0.2), we applied the Box-Cox transform to spread out the distribution, making variations more noticeable. As shown in Fig.~\ref{RQ1_sphere}, LS is highest at the reference point (0°,0°) and decreases with increasing angular distance. However, LS increases again between azimuths 120° to 180°, indicating front-back confusion. This pattern is consistent across all contents except glockenspiel and pure tone. For the glockenspiel, localization similarity increases between azimuths 60° and 90°. This is likely due to its high-frequency spectral content, which weakens ITD cues. Additionally, the pure tone heatmap shows discrete regions of high localization similarity. This could be because narrow-band, high-frequency pure tones provide weak ITD/ILD cues, leading to inconsistent localization similarity across azimuths.

Overall, the results indicate that the \mbox{BINAQUAL} is sensitive to small variations in the spatial locations of single-point sources. The consistency across different source types also demonstrates the model's robustness to variations in content.

\begin{figure*}[!t]
\centering
\includegraphics[width=\textwidth]{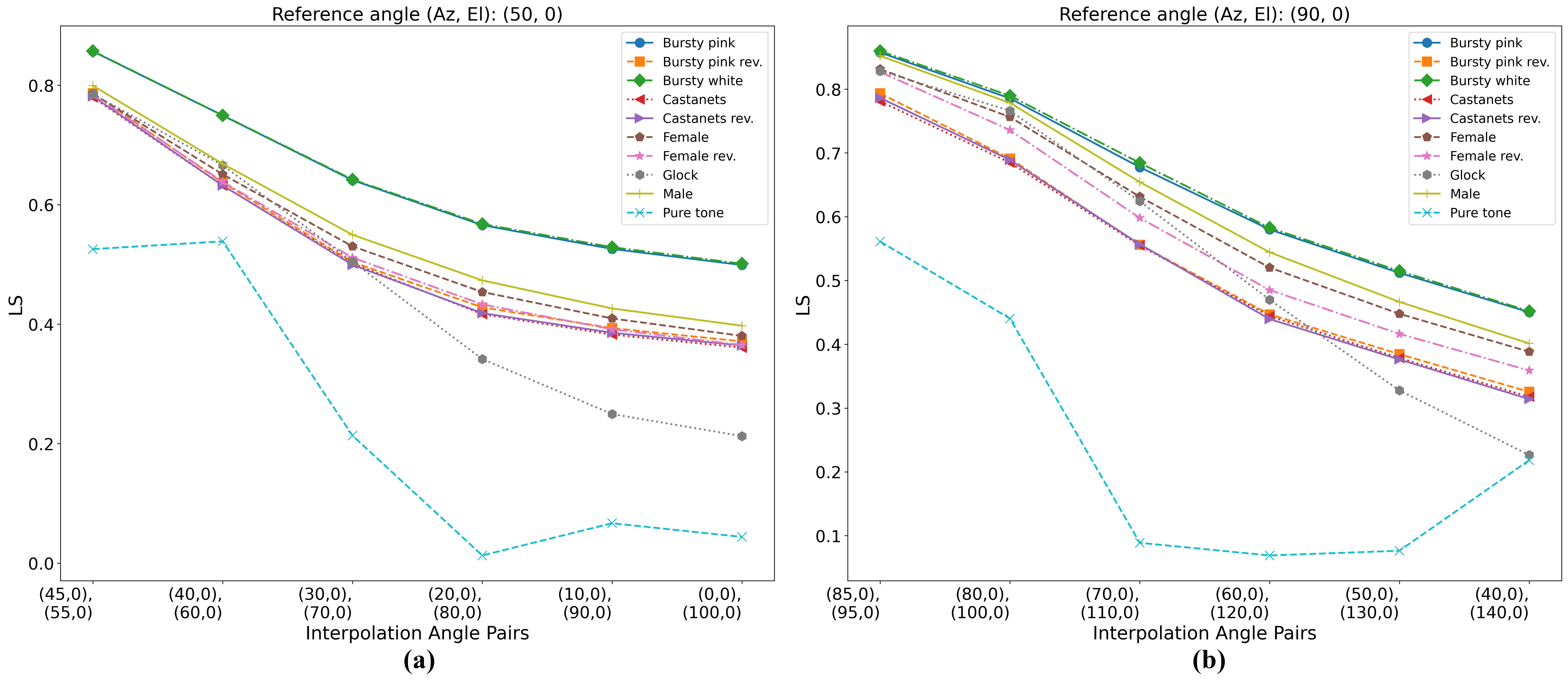}
\caption{Localization similarity between single-source interpolated test signals in the angle interpolations dataset and their corresponding references. (a) Reference is localized at azimuth = 50°, elevation = 0°. (b) Reference is localized at azimuth = 90°, elevation = 0°.}
\label{RQ3_plot}
\end{figure*}

\subsection{Angle Interpolations}\label{sec:results_interpolations}
Localization similarity between the samples in the angle interpolations dataset and their corresponding reference signals was calculated. Fig.~\ref{RQ3_plot} shows the LS across different angle interpolations for two different scenarios: (a) reference angle = (50°,0°) and (b) reference angle = (90°,0°).

As shown in Fig.~\ref{RQ3_plot}, overall, the LS decreases as interpolated angles deviate from the references. This is expected because larger angular separations introduce greater perceptual localization differences. Bursty white and bursty pink show higher LS scores across all angles. These signals are rich in broadband spectral content, which provides strong localization cues. The Pure tone signal has the lowest LS scores, indicating that it is the hardest to localize. This aligns with the known difficulty in localizing pure tones due to phase ambiguities and the lack of strong interaural cues. Glockenspiel also shows lower LS scores for more extreme interpolations. Since it has strong high-frequency content (around 1 kHz–8 kHz, with prominent overtones), the model relies more on ILD cues, which are less precise for localization. These results confirm that BIANQUAL can effectively distinguish between single-source simulated representations of different angle interpolations while being robust to variations in content.

\subsection{Surround Loudspeaker Layouts}\label{sec:results_speakers}
\begin{figure*}[!t]
\centering
\includegraphics[width=\textwidth]{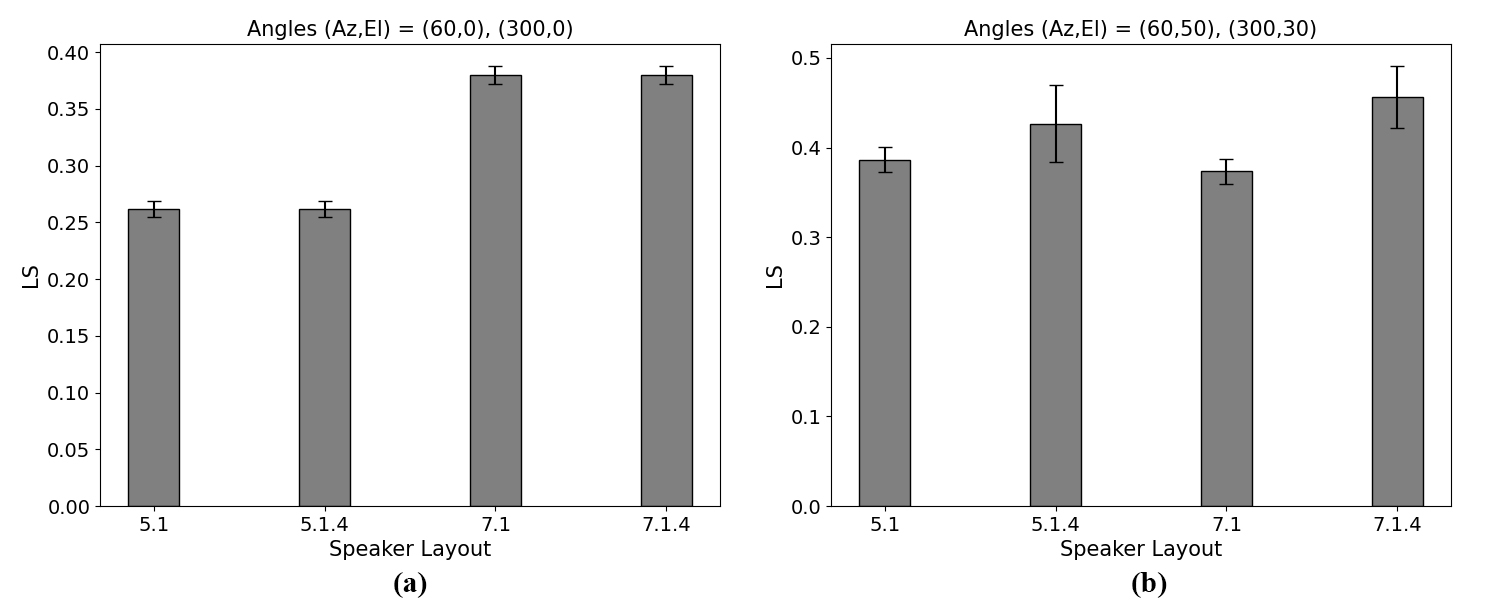}
\caption{The average localization similarity between multi-point source samples of the surround layouts dataset, rendered for different loudspeaker layouts, and corresponding reference signals without any layouts. (a) Point source angles without elevation, (b) Point source angles with elevation.}
\label{RQ2_plot}
\end{figure*}

Localization similarity between samples in the surround layouts dataset, rendered for various surround loudspeaker layouts, and reference signals (without any loudspeaker layout) was calculated. Fig.~\ref{RQ2_plot} presents the aggregated mean LS scores across different loudspeaker layouts for two scenarios: (a) source angles without elevation and (b) source angles with elevation. The mean LS scores were computed across 8 different audio samples per layout. For each configuration, the 95\% confidence interval was also calculated using the standard error of the mean across the 8 samples.

As shown in part (a) of Fig.~\ref{RQ2_plot}, when the angles have no elevation, the LS scores for 5.1 and 5.1.4 are similar, as are those for 7.1 and 7.1.4. Additionally, LS scores for 7.1 and 7.1.4 are higher than those for 5.1 and 5.1.4. This suggests that increasing the number of surround channels enhances localization similarity due to improved spatial coverage of surround loudspeakers.

As shown in part (b) of Fig.~\ref{RQ2_plot}, when angles have elevations, the LS scores for layouts with height channel (5.1.4, 7.1.4) are higher than those for layouts without elevation (5.1, 7.1). This confirms that elevated angles benefit more from overhead loudspeakers, as adding height channels allows the model to capture vertical localization cues more effectively, leading to an overall increase in LS.

Statistical analysis using non-parametric tests further supports these findings. As shown in Table~\ref{table:layout_comparison}, Mann–Whitney U tests reveal that increasing the number of surround channels leads to significantly higher LS scores, as evidenced by comparisons between 5.1 vs. 7.1 and 5.1.4 vs. 7.1.4 (both with $p = 0.0002$). Similarly, adding height channels significantly improves localization similarity for both 5.1 vs. 5.1.4 ($p = 0.0499$) and 7.1 vs. 7.1.4 ($p = 0.0003$). Kruskal–Wallis tests across all layouts further confirm overall significant differences due to layout configuration. These results demonstrate that \mbox{BINAQUAL} can distinguish between simulated binaural representations of different surround loudspeaker layouts, regardless of whether elevation is present.

\begin{figure*}[!t]
\centering
\includegraphics[width=\textwidth]{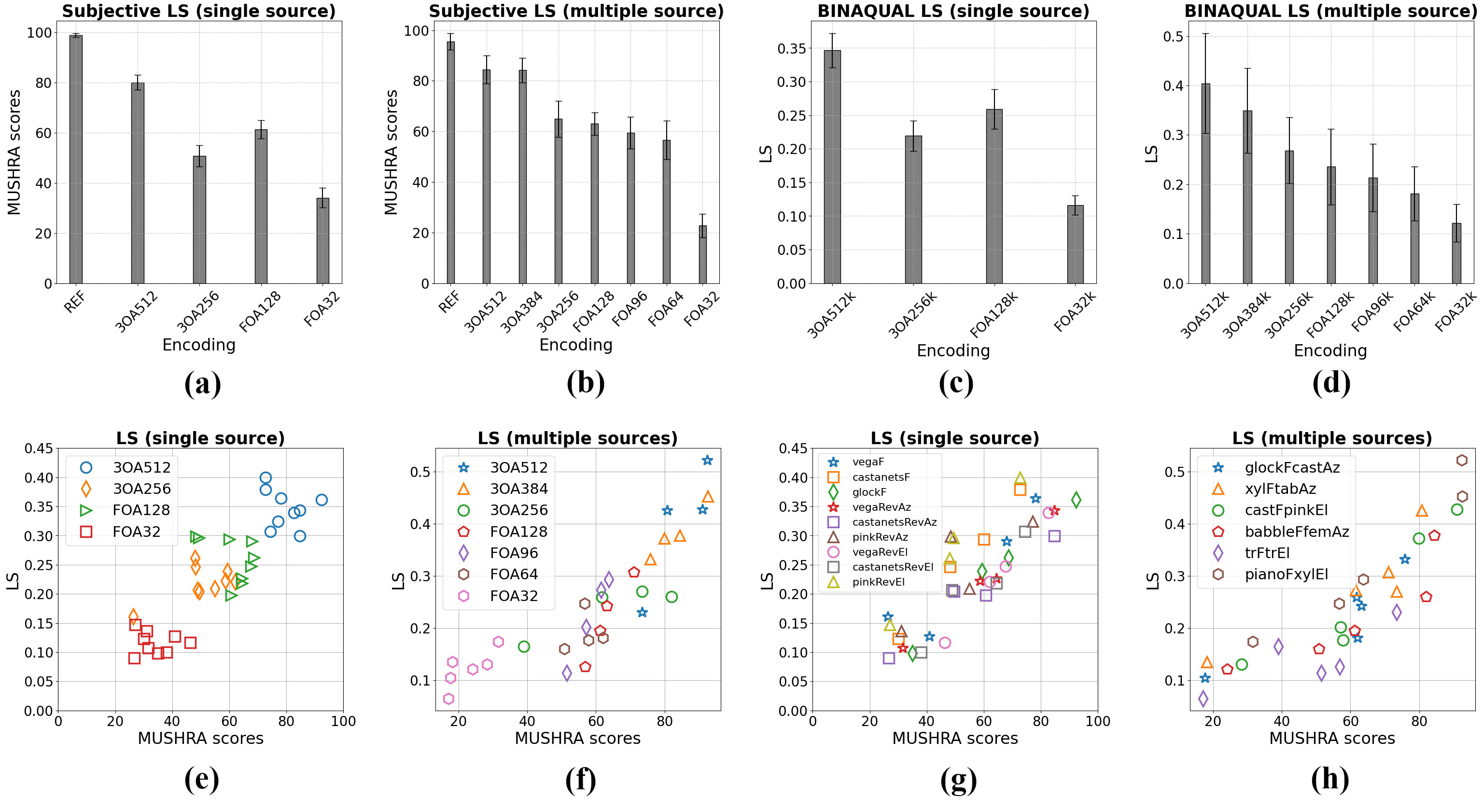}
\caption{Subjective localization similarity vs. \mbox{BINAQUAL} results: Plots (a) and (b) show the aggregated subjective MUSHRA test results for different encodings in single-point and multiple-point source scenarios, respectively. Plots (c) and (d) present the aggregated \mbox{BINAQUAL} LS scores for different encodings in single-point and multiple-point source scenarios, respectively. Plots (e) and (f) illustrate the relationship between \mbox{BINAQUAL} LS and MUSHRA scores, grouped by encoding, for single-point and multiple-point source scenarios, respectively. Plots (g) and (h) show the relationship between \mbox{BINAQUAL} LS and MUSHRA scores, grouped by audio content, for single-point and multiple-point source scenarios, respectively.}
\label{RQ4_plot}
\end{figure*}

\subsection{Codec Compression}\label{sec:results_compression}
Localization similarity was computed between all single-point and multiple-point samples of the codec compression dataset and their corresponding references (non-compressed signals). Fig. 6 compares the aggregated LS scores with the results of a subjective listening test performed in \cite{Narbutt2020AMBIQUAL} with the same set of recordings.

Fig.~\ref{RQ4_plot} (a) and (b) show the aggregated mean values of subjective MUSHRA results across various encodings for single-point and multiple-point source scenarios, respectively. Parts (c) and (d) of Fig.~\ref{RQ4_plot} present the \mbox{BINAQUAL} LS results for different encodings in single-source and multiple-source scenarios, respectively. The 95\% confidence intervals are also shown. We see similar trends between aggregated subjective results (MUSHRA) and aggregated \mbox{BINAQUAL} LS scores. This holds for both single-point and multi-point source scenarios across different audio contents.

Fig.~\ref{RQ4_plot} (e) and Fig.~\ref{RQ4_plot} (f) depict the relationship between \mbox{BINAQUAL} LS scores and subjective MUSHRA results across different encodings, while Fig.~\ref{RQ4_plot} (g) and Fig.~\ref{RQ4_plot} (h) illustrate similar comparisons for different audio contents for both single-point and multiple-point source scenarios. We observe that the data points have been clustered by encoding, while at the same time, there is a lack of clustering by content.

There is a strong correlation between subjective MUSHRA results and LS scores. For single-point source results, the Pearson and Spearman correlations are 0.85 and 0.84, respectively. For multiple-point source results, these correlations increase to 0.87 and 0.92, respectively, suggesting a slightly stronger monotonic relationship in the multiple-point scenario. 

Additionally, we computed the correlation between \mbox{AMBIQUAL} localization accuracy scores and \mbox{BINAQUAL} LS scores. For single-point sources, the Pearson and Spearman correlation coefficients are 0.84 and 0.80, respectively. For multiple-point sources, the correlations increase to 0.93 (Pearson) and 0.96 (Spearman).

The results indicate that the BINAQAUL model effectively distinguishes between different compression levels of binaural audio signals while remaining robust to variations in content.





\subsection{General Discussion}
The results presented in this study demonstrate that the \mbox{BINAQUAL} is effective in assessing spatial localization similarity in binaural audio across various conditions. In Sec. \ref{sec:results_ls}, we analyzed the model's sensitivity to source localization and showed that \mbox{BINAQUAL} successfully detects small variations in point-source locations, as evidenced by the decrease in LS scores with increasing angular distance from the reference. However, for angles in the range of ±120° to ±180°, LS scores increased, which could be attributed to front-back confusion where binaural cues are ambiguous due to the lack of sufficient spectral differentiation. This observation aligns with psychoacoustic phenomena in binaural hearing \cite{Blauert1997Spatial}. Additionally, we found that the model is less effective in localizing pure tones due to their lack of rich interaural cues, a limitation that is consistent with human difficulties in localizing narrow-band audio sources \cite{Moore2012introduction}. The results also indicate that \mbox{BINAQUAL} maintains relatively flat LS scores at extreme elevations (±90°). This behavior is expected, as azimuth resolution effectively collapses at ±90° elevation. Also, elevation perception relies more heavily on spectral cues, which are typically more variable and harder to model robustly, further contributing to the observed uniformity in LS scores.

In Sec. \ref{sec:results_interpolations}, we analyzed the model’s ability to differentiate between angle interpolations of single-point sources and found that \mbox{BINAQUAL} tracks the decrease in LS as the azimuth angles deviate from the reference position. We saw a similar trend across all contents except pure tone, which again confirms the difficulty of localizing narrow-band signals due to the lack of interaural cues. Similarly, glockenspiel, which has high-frequency spectral content, demonstrated lower LS for more extreme angles, suggesting that a lack of strong ITD cues can reduce localization precision.

Sec. \ref{sec:results_speakers} showed that \mbox{BINAQUAL} could differentiate effectively between 5.1-, 5.1.4-, 7.1-, and 7.1.4-rendered audio signals, showing that surround loudspeaker layouts with more channels (and especially additional height channels) increase localization similarity. The capability to discriminate among different loudspeaker layouts is particularly promising for applications involving complex multichannel setups in VR, AR, and cinematic audio.

Sec. \ref{sec:results_compression} explored the model’s ability to distinguish between different compression levels of binaural audio content. \mbox{BINAQUAL} not only differentiated among binaural renders of different compression levels but also correlated strongly with subjective MUSHRA tests. The model achieved higher correlation coefficients in multi-source scenarios, indicating that \mbox{BINAQUAL} generalizes well to complex scenes containing multiple sound sources.

Finally, regarding its robustness to variations in content and the number of sources, the model demonstrated stable performance across both single- and multi-point source recordings, as well as different audio types (e.g., speech, music, impulsive sounds). This versatility indicates that \mbox{BINAQUAL} is applicable to a wide range of practical scenarios, from simple point-source recordings to more complex, multi-source mixes.

While \mbox{BINAQUAL} effectively generalizes across content, source configurations, and degradations, some limitations remain. Its performance declines in conditions where binaural cues are inherently ambiguous or diminished, such as at extreme elevations, with pure tone stimuli, or in cases of front-back confusion. These are also well-known challenges for human listeners, which reinforces the perceptual validity of the metric. We should note that our goal is not to outperform human hearing, but rather to develop a metric that closely mimics how humans perceive spatial audio quality. Future work could enhance \mbox{BINAQUAL} by integrating additional spectral cues or simulating listener head movements, strategies that humans naturally use to resolve spatial ambiguities. Such enhancements aim to further align the metric with perceptual realism and improve its applicability to dynamic or immersive spatial audio scenarios.

We addressed all five research questions outlined in Section \ref{sec:intro}, and the results demonstrated the capabilities of the \mbox{BINAQUAL} model. While most existing spatial localization quality metrics focus primarily on point-source localization accuracy, we took a broader approach to spatial audio evaluation. By examining the effects of angle interpolations, loudspeaker layouts, codec compression, and content variations, our work provides deeper insights into how spatial fidelity is preserved across diverse scenarios. These aspects are crucial for advancing localization quality assessment models and ensuring objective metrics align more closely with perceptual realism. Consequently, this study not only introduces a novel localization quality assessment metric but also sets a foundation for a more holistic approach to spatial localization quality evaluation beyond traditional accuracy-based methods.

\section{CONCLUSION}\label{sec:conclusion}
This study presented a full-reference objective localization similarity metric for binaural audio. Through a comprehensive set of validation experiments, we demonstrated the model’s capabilities in differentiating between variations in source positions, angle interpolations, surround loudspeaker layouts, and audio degradations. 

The model successfully differentiates subtle spatial variations, as evidenced by its sensitivity to small angular deviations. However, its performance is affected by well-known psychoacoustic limitations, such as front-back confusion and the reduced localization accuracy of pure tones. These findings align with established auditory localization principles, reinforcing the model's validity. Also, \mbox{BINAQUAL} effectively distinguishes between different surround loudspeaker layouts, highlighting its potential for evaluating complex binaural audio formats.

Our analysis of compressed binaural audio demonstrates \mbox{BINAQUAL}’s strong correlation with perceptual MUSHRA scores, particularly in multi-source scenarios. This indicates that the metric can be a reliable proxy for subjective evaluations, offering a scalable and efficient alternative to listener-based assessments. Furthermore, \mbox{BINAQUAL} maintains robust performance across different content types and varying numbers of sound sources, indicating its applicability across diverse spatial audio scenarios.

Our work expands spatial audio quality evaluation beyond point-source localization accuracy by considering factors like angle interpolations, loudspeaker layouts, compression artifacts, and content variations, offering a more comprehensive assessment of spatial fidelity in binaural audio. Future work could further refine \mbox{BINAQUAL}'s performance to ensure even greater alignment with human spatial perception.


\section{Acknowledgment}
This publication has emanated from research conducted with the financial support of Taighde Éireann – Research Ireland under Grant numbers 12/RC/2289\_P2 and 13/RC/2077\_P2 and via a gift to support this research from Google. For the purpose of Open Access, the author has applied a CC BY public copyright license to any Author Accepted Manuscript version arising from this submission.

\bibliography{refs}

\begin{thebibliography}{10}

\bibitem{Pangarkar2025Smart}
Tajammul Pangarkar.
\newblock Smart {Wearables} {Statistics} and {Facts} (2025).
\newblock https://scoop.market.us/smart-wearables-statistics/, jan 14 2025.
\newblock [Online; accessed 2025-02-18].

\bibitem{eclipsa_audio}
Introducing {Eclipsa} {Audio}: immersive audio for everyone {\textbar} {Google} {Open} {Source} {Blog}.

\bibitem{RECOMMENDATION}
Recommendation {ITU}-{R} {BS}.1534-1 - {Method} for the subjective assessment of intermediate quality level of coding systems.
\newblock {\em R BS.}, page~18.

\bibitem{Makous1990Twodimensional}
James~C. Makous and John~C. Middlebrooks.
\newblock Two dimensional sound localization by human listeners.
\newblock {\em The Journal of the Acoustical Society of America}, 87(5):2188--2200, may 1 1990.

\bibitem{mills1958minimum}
Allen~William Mills.
\newblock On the minimum audible angle.
\newblock {\em The Journal of the Acoustical Society of America}, 30(4):237--246, 1958.

\bibitem{Bech2007Perceptual}
S\o{}ren Bech and Nick Zacharov.
\newblock {\em Perceptual audio evaluation-{Theory}, method and application}.
\newblock John Wiley \& Sons, 2007.

\bibitem{Ihlefeld2011Effect}
Antje Ihlefeld and Barbara~G. Shinn-Cunningham.
\newblock Effect of source spectrum on sound localization in an everyday reverberant room.
\newblock {\em The Journal of the Acoustical Society of America}, 130(1):324--333, jul 19 2011.

\bibitem{Choi2007Prediction}
In~Yong Choi, Sang~Bae Chon, Barbara~G Shinn-Cunningham, and Koeng-Mo Sung.
\newblock Prediction of perceived quality in multi-channel audio compression coding systems.
\newblock In {\em Audio Engineering Society Conference: 30th International Conference: Intelligent Audio Environments}. Audio Engineering Society, 2007.

\bibitem{Flessner2017Assessment}
Jen-Hendrik Flessner, Rainer Huber, and Stephan Ewert.
\newblock Assessment and {Prediction} of {Binaural} {Aspects} of {Audio} {Quality}.
\newblock {\em Journal of the Audio Engineering Society}, 65(11):929--942, nov 28 2017.

\bibitem{Flener2019Subjective}
Jan-Hendrik Fle\ss{}ner, Thomas Biberger, and Stephan~D. Ewert.
\newblock Subjective and {Objective} {Assessment} of {Monaural} and {Binaural} {Aspects} of {Audio} {Quality}.
\newblock {\em IEEE/ACM Transactions on Audio, Speech, and Language Processing}, 27(7):1112--1125, 7 2019.
\newblock event-title: IEEE/ACM Transactions on Audio, Speech, and Language Processing.

\bibitem{Watcharasupat2023Quantifying}
Karn~N Watcharasupat and Alexander Lerch.
\newblock Quantifying spatial audio quality impairment.
\newblock In {\em ICASSP 2024-2024 IEEE International Conference on Acoustics, Speech and Signal Processing (ICASSP)}, pages 746--750. IEEE, 2024.

\bibitem{Flesner2014Quality}
Jan-Hendrik Flesner, Stephan~D. Ewert, Birger Kollmeier, and Rainer Huber.
\newblock Quality assessment of multi-channel audio processing schemes based on a binaural auditory model.
\newblock In {\em 2014 {IEEE} {International} {Conference} on {Acoustics}, {Speech} and {Signal} {Processing} ({ICASSP})}, pages 1340--1344, Florence, Italy, 5 2014. IEEE.

\bibitem{Manocha2021DPLM}
Pranay Manocha, Anurag Kumar, Buye Xu, Anjali Menon, Israel~D. Gebru, Vamsi~K. Ithapu, and Paul Calamia.
\newblock Dplm: A {Deep} {Perceptual} {Spatial}-{Audio} {Localization} {Metric}.
\newblock In {\em 2021 {IEEE} {Workshop} on {Applications} of {Signal} {Processing} to {Audio} and {Acoustics} ({WASPAA})}, pages 6--10, 10 2021.
\newblock ISSN: 1947-1629.

\bibitem{Manocha2022SAQAM}
Pranay Manocha, Anurag Kumar, Buye Xu, Anjali Menon, Israel~Degene Gebru, Vamsi~Krishna Ithapu, and Paul Calamia.
\newblock Saqam: Spatial {Audio} {Quality} {Assessment} {Metric}.
\newblock In {\em Interspeech 2022}, pages 649--653. ISCA, sep 18 2022.

\bibitem{Manocha2023Spatialization}
Pranay Manocha, Israel~Dejene Gebru, Anurag Kumar, Dejan Markovic, and Alexander Richard.
\newblock Spatialization {Quality} {Metric} for {Binaural} {Speech}.
\newblock In {\em INTERSPEECH 2023}, pages 5426--5430. ISCA, aug 20 2023.

\bibitem{Narbutt2020AMBIQUAL}
Miroslaw Narbutt, Jan Skoglund, Andrew Allen, Michael Chinen, Dan Barry, and Andrew Hines.
\newblock Ambiqual: Towards a {Quality} {Metric} for {Headphone} {Rendered} {Compressed} {Ambisonic} {Spatial} {Audio}.
\newblock {\em Applied Sciences}, 10(9):3188, 1 2020.
\newblock number: 9 publisher: Multidisciplinary Digital Publishing Institute.

\bibitem{Mller1992Fundamentals}
Henrik M\o{}ller.
\newblock Fundamentals of binaural technology.
\newblock {\em Applied Acoustics}, 36(3):171--218, jan 1 1992.

\bibitem{Moore2010Oxford}
David~R Moore, Paul~Albert Fuchs, Adrian Rees, Alan Palmer, and Christopher~J Plack.
\newblock {\em The {Oxford} handbook of auditory science: The auditory brain}, volume~2.
\newblock Oxford University Press, USA, 2010.

\bibitem{Macpherson2002Listener}
Ewan~A. Macpherson and John~C. Middlebrooks.
\newblock Listener weighting of cues for lateral angle: the duplex theory of sound localization revisited.
\newblock {\em The Journal of the Acoustical Society of America}, 111(5 Pt 1):2219--2236, 5 2002.
\newblock PMID: 12051442.

\bibitem{jc1991sound}
Middlebrooks Jc.
\newblock Sound localization by human listeners.
\newblock {\em Ann Rev Psychol}, 42:135--159, 1991.

\bibitem{Blauert1997Spatial}
Jens Blauert.
\newblock {\em Spatial {Hearing}: The {Psychophysics} of {Human} {Sound} {Localization}}.
\newblock MIT Press, 1997.
\newblock Google-Books-ID: wBiEKPhw7r0C.

\bibitem{Begault20003}
Durand~R Begault and Leonard~J Trejo.
\newblock 3-{D} sound for virtual reality and multimedia.
\newblock Technical report, 2000.

\bibitem{Armstrong2018Perceptual}
Cal Armstrong, Lewis Thresh, Damian Murphy, and Gavin Kearney.
\newblock A {Perceptual} {Evaluation} of {Individual} and {Non}-{Individual} {HRTFs}: A {Case} {Study} of the {SADIE} {II} {Database}.
\newblock {\em Applied Sciences}, 8(11):2029, 11 2018.
\newblock number: 11 publisher: Multidisciplinary Digital Publishing Institute.

\bibitem{dobreva2011influence}
Marina~S Dobreva, William~E O'Neill, and Gary~D Paige.
\newblock Influence of aging on human sound localization.
\newblock {\em Journal of neurophysiology}, 105(5):2471--2486, 2011.

\bibitem{Hofman1998Relearning}
Paul~M. Hofman, Jos G.~A. Van~Riswick, and A.~John Van~Opstal.
\newblock Relearning sound localization with new ears.
\newblock {\em Nature Neuroscience}, 1(5):417--421, 9 1998.
\newblock publisher: Nature Publishing Group.

\bibitem{Brungart1999Auditory}
Douglas~S. Brungart and William~M. Rabinowitz.
\newblock Auditory localization of nearby sources. {Head}-related transfer functions.
\newblock {\em The Journal of the Acoustical Society of America}, 106(3):1465--1479, sep 1 1999.

\bibitem{Wightman1999Resolution}
F.~L. Wightman and D.~J. Kistler.
\newblock Resolution of front-back ambiguity in spatial hearing by listener and source movement.
\newblock {\em The Journal of the Acoustical Society of America}, 105(5):2841--2853, 5 1999.
\newblock PMID: 10335634.

\bibitem{Hartmann1983Localization}
W.~M. Hartmann.
\newblock Localization of sound in rooms.
\newblock {\em The Journal of the Acoustical Society of America}, 74(5):1380--1391, nov 1 1983.

\bibitem{Kerber2012Sound}
Stefan Kerber and Bernhard~U. Seeber.
\newblock Sound localization in noise by normal-hearing listeners and cochlear implant users.
\newblock {\em Ear and hearing}, 33(4):445--457, 2012.
\newblock PMID: 22588270 PMCID: PMC3446659.

\bibitem{Ege2018Accuracy}
Rachel Ege, A.~John~Van Opstal, and Marc~M. Van~Wanrooij.
\newblock Accuracy-{Precision} {Trade}-off in {Human} {Sound} {Localisation}.
\newblock {\em Scientific Reports}, 8(1):16399, nov 6 2018.
\newblock publisher: Nature Publishing Group.

\bibitem{Shigeno1983Localization}
Sumi Shigeno and Tadasu Oyama.
\newblock Localization of speech and non-speech sounds.
\newblock {\em Japanese Psychological Research}, 25(2):112--117, 1983.
\newblock publisher-place: United Kingdom publisher: Blackwell Publishing.

\bibitem{Chau1995Combined}
W.~Chau and R.O. Duda.
\newblock Combined monaural and binaural localization of sound sources.
\newblock In {\em Conference {Record} of {The} {Twenty}-{Ninth} {Asilomar} {Conference} on {Signals}, {Systems} and {Computers}}, volume~2, pages 1281--1285 vol.2, 10 1995.
\newblock ISSN: 1058-6393.

\bibitem{Borg2008Effect}
Erik Borg, Christina Bergkvist, and Dan Bagger-Sj{\" o}b{\" a}ck.
\newblock Effect on {Directional} {Hearing} in {Hunters} {Using} {Amplifying} ({Level} {Dependent}) {Hearing} {Protectors}.
\newblock {\em Otology \& Neurotology}, 29(5):579, 8 2008.

\bibitem{Middlebrooks1992Narrow}
J.~C. Middlebrooks.
\newblock Narrow-band sound localization related to external ear acoustics.
\newblock {\em The Journal of the Acoustical Society of America}, 92(5):2607--2624, 11 1992.
\newblock PMID: 1479124.

\bibitem{Butler1992Localization}
R.~A. Butler and R.~A. Humanski.
\newblock Localization of sound in the vertical plane with and without high-frequency spectral cues.
\newblock {\em Perception \& Psychophysics}, 51(2):182--186, 2 1992.
\newblock PMID: 1549436.

\bibitem{Wightman1992dominant}
F.~L. Wightman and D.~J. Kistler.
\newblock The dominant role of low-frequency interaural time differences in sound localization.
\newblock {\em The Journal of the Acoustical Society of America}, 91(3):1648--1661, 3 1992.
\newblock PMID: 1564201.

\bibitem{Yost2014Sound}
William~A. Yost and Xuan Zhong.
\newblock Sound source localization identification accuracy: Bandwidth dependencies.
\newblock {\em The Journal of the Acoustical Society of America}, 136(5):2737--2746, nov 1 2014.

\bibitem{zakarauskas1993computational}
Pierre Zakarauskas and Max~S Cynader.
\newblock A computational theory of spectral cue localization.
\newblock {\em The Journal of the Acoustical Society of America}, 94(3):1323--1331, 1993.

\bibitem{langendijk2002contribution}
Erno~HA Langendijk and Adelbert~W Bronkhorst.
\newblock Contribution of spectral cues to human sound localization.
\newblock {\em The Journal of the Acoustical Society of America}, 112(4):1583--1596, 2002.

\bibitem{dietz2011auditory}
Mathias Dietz, Stephan~D Ewert, and Volker Hohmann.
\newblock Auditory model based direction estimation of concurrent speakers from binaural signals.
\newblock {\em Speech Communication}, 53(5):592--605, 2011.

\bibitem{may2010probabilistic}
Tobias May, Steven Van De~Par, and Armin Kohlrausch.
\newblock A probabilistic model for robust localization based on a binaural auditory front-end.
\newblock {\em IEEE Transactions on Audio, Speech, and Language Processing}, 19(1):1--13, 2010.

\bibitem{lindemann1986extension}
Werner Lindemann.
\newblock Extension of a binaural cross-correlation model by contralateral inhibition. i. simulation of lateralization for stationary signals.
\newblock {\em The Journal of the Acoustical Society of America}, 80(6):1608--1622, 1986.

\bibitem{baumgartner2014modeling}
Robert Baumgartner, Piotr Majdak, and Bernhard Laback.
\newblock Modeling sound-source localization in sagittal planes for human listeners.
\newblock {\em The Journal of the Acoustical Society of America}, 136(2):791--802, 2014.

\bibitem{ren2025metric}
Hualin Ren, Christian Ritz, Jiahong Zhao, Xiguang Zheng, and Daeyoung Jang.
\newblock A metric for predicting the quality of ambisonic spatial audio reproduced using spatially interpolated or extrapolated room impulse responses.
\newblock In {\em ICASSP 2025-2025 IEEE International Conference on Acoustics, Speech and Signal Processing (ICASSP)}, pages 1--5. IEEE, 2025.

\bibitem{zheng2025hapg}
Yuanming Zheng, Jiaxuan Yao, Xiangyu Deng, Yuhong Yang, Ruiqi Liao, Weiping Tu, and Cedar Lin.
\newblock Hapg-saqam: Human auditory perception guided spatial audio quality assessment metric.
\newblock In {\em ICASSP 2025-2025 IEEE International Conference on Acoustics, Speech and Signal Processing (ICASSP)}, pages 1--5. IEEE, 2025.

\bibitem{Hines2015ViSQOLAudio}
Andrew Hines, Eoin Gillen, Damien Kelly, Jan Skoglund, Anil Kokaram, and Naomi Harte.
\newblock Visqolaudio: An objective audio quality metric for low bitrate codecs.
\newblock {\em The Journal of the Acoustical Society of America}, 137(6):EL449--EL455, may 28 2015.

\bibitem{Hines2012Speech}
Andrew Hines and Naomi Harte.
\newblock Speech intelligibility prediction using a {Neurogram} {Similarity} {Index} {Measure}.
\newblock {\em Speech Communication}, 54(2):306--320, 2 2012.

\bibitem{binamix}
Dan Barry, Davoud Shariat~Panah, Alessandro Ragano, Jan Skoglund, and Andrew Hines.
\newblock Binamix - a python library for generating binaural audio datasets.
\newblock In {\em Audio Engineering Society Convention 158}. Audio Engineering Society, 2025.

\bibitem{Tech19883253}
EBU Tech.
\newblock 3253-{E}, {Sound} quality assessment material,'' {SQUAM} {CD} ({Handbook}).
\newblock {\em EBU Technical Centre Brussels}, 1988.

\bibitem{pulkki1997virtual}
Ville Pulkki.
\newblock Virtual sound source positioning using vector base amplitude panning.
\newblock {\em Journal of the Audio Engineering Society}, 45(6):456--466, 1997.

\bibitem{Moore2012introduction}
Brian~CJ Moore.
\newblock {\em An introduction to the psychology of hearing}.
\newblock Brill, 2012.

\end{thebibliography}
\bibliographystyle{unsrt} 


\clearpage  
\appendix

\section*{APPENDIX}
\renewcommand{\thetable}{A\arabic{table}}
\setcounter{table}{0}

\begin{table*}[h]
    \centering
    \tabcolsep8.1pt
    \caption{The increments for different azimuth ranges. These values were determined by calculating the average human localization error across different elevations for the following azimuth ranges.}
    \label{table:azimuth_increments}
    {%
    \begin{tabular}{@{}rrS@{}}
        \toprule
        \textbf{From (degree)} & \textbf{To (degree)} & \textbf{Increment (degree)} \\\midrule
        0   & 59   & 5  \\
        60  & 119  & 8  \\
        120 & 179  & 12 \\
        180 & 239  & 12 \\
        240 & 299  & 8  \\
        300 & 359  & 5  \\\bottomrule
    \end{tabular}}
\end{table*}

\begin{table*}[h]
    \centering
    \tabcolsep8.1pt
    \caption{The increments for different elevation ranges. These values were determined by calculating the average human localization error across different azimuths for the following elevation ranges.}
    \label{table:elevation_increments}
    {%
    \begin{tabular}{@{}rrS@{}}
        \toprule
        \textbf{From (degree)} & \textbf{To (degree)} & \textbf{Increment (degree)} \\\midrule
        \hphantom{-}0   & \hphantom{-}29   & 5  \\
        \hphantom{-}30  & \hphantom{-}59   & 10 \\
        \hphantom{-}60  & \hphantom{-}90   & 15 \\
        \hphantom{-}0   & -29  & 5  \\
        -30 & -59  & 10 \\
        -60 & -90  & 15 \\\bottomrule
    \end{tabular}}
\end{table*}

\begin{table*}[h]
    \centering
    \tabcolsep8.1pt
    \caption{Description of the base mono audio samples used for synthesizing the localization sensitivity dataset.}
    \label{table:audio_content}
    {%
    \begin{tabular}{@{}llll@{}}
        \toprule
        \textbf{Audio content} & \textbf{Frequency} & \textbf{Bandwidth} & \textbf{Source} \\\midrule
         Bursty white noise            & Low, Mid, High & Broad & Synthetic \\
        Bursty pink noise            & Low, Mid    & Broad   & Synthetic \\
        Bursty pink noise w. reverb   & Low, Mid    & Broad   & Synthetic \\
        Male speech                   & Low, Mid    & Medium  & EBU \\
        Female speech                 & Mid, High   & Medium  & EBU \\
        Female speech w. reverb       & Mid, High   & Medium  & Processed EBU \\
        Castanets                      & Mid, High  & Medium  & EBU \\
        Castanets w. reverb            & Mid, High  & Medium  & Processed EBU \\
         Glockenspiel                   & High       & Medium  & EBU \\
        Pure tone (4 kHz)              & High       & Narrow  & Synthetic \\\bottomrule
    \end{tabular}}
\end{table*}

\begin{table*}[h]
    \centering
    \tabcolsep8.1pt
    \caption{Description of the two-point sources used for synthesizing the surround layouts dataset. The last column summarizes the testing purpose of each combination.}
    \label{table:rq2}
    {%
    \begin{tabular}{@{}p{4cm} p{4cm} p{7cm}@{}}
        \toprule
        \textbf{Source \#1} & \textbf{Source \#2} & \textbf{Testing purpose} \\\midrule
        Male speech                 & Female speech w. reverb     & Localizing two vocal sources with different frequency ranges \\
        Castanets                  & Male speech                & Localizing a percussive natural sound against a vocal source \\
        Bursty pink noise          & Female speech              & Localizing a synthetic broadband sound against a vocal source \\
        Bursty white noise         & Male speech                & Localizing a synthetic broadband sound against a vocal source \\
        Bursty pink noise w. reverb & Female speech w. reverb    & Localizing two reverberant sounds with different characteristics \\
        Bursty white noise         & Bursty pink noise w. reverb & Localizing two synthetic broadband sounds with distinct temporal properties \\
        Bursty pink noise          & Castanets         & Localizing a synthetic sound against a natural percussive sound \\
        Bursty white noise         & Castanets w. reverb         & Localizing a synthetic sound against a reverberant percussive natural sound \\\bottomrule

    \end{tabular}}
\end{table*}

\begin{table*}[]
    \centering
    \tabcolsep8.1pt
    \caption{The angle pairs used to generate the angle interpolations dataset.}
    \label{table:rq3}
    {%
    \begin{tabular}{@{}cc|cc@{}}
        \toprule
        \multicolumn{4}{c}{\textbf{Reference: Azimuth = 50°, Elevation = 0°}} \\\midrule
        \textbf{Azimuth 1} & \textbf{Elevation 1} & \textbf{Azimuth 2} & \textbf{Elevation 2} \\\midrule
        45°  & 0°  & 55°  & 0°  \\
        40°  & 0°  & 60°  & 0°  \\
        30°  & 0°  & 70°  & 0°  \\
        20°  & 0°  & 80°  & 0°  \\
        10°  & 0°  & 90°  & 0°  \\
        0°   & 0°  & 100° & 0°  \\\bottomrule
    \end{tabular}}

    \vspace{10pt} 

    {%
    \begin{tabular}{@{}cc|cc@{}}
        \toprule
        \multicolumn{4}{c}{\textbf{Reference: Azimuth = 90°, Elevation = 0°}} \\\midrule
        \textbf{Azimuth 1} & \textbf{Elevation 1} & \textbf{Azimuth 2} & \textbf{Elevation 2} \\\midrule
        85°  & 0°  & 95°  & 0°  \\
        80°  & 0°  & 100° & 0°  \\
        70°  & 0°  & 110° & 0°  \\
        60°  & 0°  & 120° & 0°  \\
        50°  & 0°  & 130° & 0°  \\
        40°  & 0°  & 140° & 0°  \\\bottomrule
    \end{tabular}}
\end{table*}

\begin{table*}[]
    \centering
    \tabcolsep8.1pt
    \caption{Statistical comparison between different surround loudspeaker layouts using Mann–Whitney and Kruskal–Wallis tests.}
    \label{table:layout_comparison}

     {%
    \begin{tabular}{@{}lcc@{}}
        \toprule
        \multicolumn{3}{c}{\textbf{Effect of increasing the layout size}} \\\midrule
        \textbf{Comparison} & \textbf{Test} & \textbf{p-value} \\\midrule
        5.1 vs 7.1 & Mann–Whitney U & 0.0002 \\
        5.1.4 vs 7.1.4 & Mann–Whitney U & 0.0002 \\
        All layouts & Kruskal–Wallis & 0.0000 \\\bottomrule
    \end{tabular}}

        \vspace{10pt}

    {%
    \begin{tabular}{@{}lcc@{}}
        \toprule
        \multicolumn{3}{c}{\textbf{Effect of adding more height channels}} \\\midrule
        \textbf{Comparison} & \textbf{Test} & \textbf{p-value} \\\midrule
        5.1 vs 5.1.4 & Mann–Whitney U & 0.0499 \\
        7.1 vs 7.1.4 & Mann–Whitney U & 0.0003 \\
        All layouts & Kruskal–Wallis & 0.0008 \\\bottomrule
    \end{tabular}}

\end{table*}

\clearpage

\end{document}